\documentclass{ws-procs975x65}

\def\Msun{\hbox{$\rm\thinspace M_{\odot}$}}

\begin{document}

\title{Atomic X-ray Spectra of Accretion Disk Atmospheres in the Kerr Metric}

\author{MARIO A. JIMENEZ-GARATE
\footnote{\uppercase{I} acknowledge support from \uppercase{NASA}
contract \uppercase{NAS 8-01129}.}}

\address{Massachusetts Institute of Technology,
Center for Space Research, 70 Vassar St, NE80-6009,
Cambridge, MA, 02139, USA.  E-mail: mario@alum.mit.edu}

\author{DUANE A. LIEDAHL, 
CHRISTOPHER W. MAUCHE\footnote{\uppercase{T}his work was 
performed under the auspices of the \uppercase{U.S.}\ \uppercase{D}epartment 
of \uppercase{E}nergy by 
\uppercase{U}niversity of \uppercase{C}alifornia \uppercase{L}awrence 
\uppercase{L}ivermore \uppercase{N}ational \uppercase{L}aboratory under 
contract \uppercase{N}o.\ \uppercase{W}-7405-\uppercase{E}ng-48.} }

\address{Lawrence Livermore National Laboratory, Livermore, CA, 94550, USA.}

\author{JOHN C. RAYMOND}

\address{Harvard-Smithsonian Center for Astrophysics, Cambridge, MA, 02138, USA.}

\maketitle

\abstracts{We calculate the atmospheric structure of an accretion disk
around a Kerr black hole and obtain its X-ray spectrum, 
which exhibits prominent atomic transitions under certain circumstances.
The gravitational and Doppler (red)shifts
of the C~V, C~VI, O~VII, O~VIII, and Fe~I--XXVI 
emission lines are observable in active galaxies.  We quantify the 
line emissivities as a function of radius, to 
identify the effects of atmospheric structure, and to
determine the usefulness of these lines for probing the disk energetics. 
The line emissivities do not always scale linearly with the incident 
radiative energy, as in the case of Fe~XXV and Fe~XXVI.
Our model incorporates photoionization and thermal balance for the plasma, 
the hydrostatic approximation perpendicular to the plane of the disk,
and general relativistic tidal forces.
We include radiative recombination rates, fluorescence yields, 
Compton scattering, and photoelectric opacities for the most abundant elements.}

\section{Introduction}
One of the key signatures of strong gravity in astrophysical black holes
is the peculiar shape of the Fe K$\alpha$ emission line observed
at 6.4~keV in active galactic nuclei and black hole X-ray binaries.\cite{rev}
The Fe K$\alpha$ emission has been extensively
studied in model calculations and Monte Carlo radiation transfer models.  
In such models, the Fe K$\alpha$ line emission is 
produced by near-neutral plasma 
in the accretion disk that is irradiated by an external source of nonthermal 
X-rays and $\gamma$-rays,
located somewhere above the accretion disk.
The Fe K$\alpha$ line is skewed and redshifted down to $\sim 3$ keV, but it only 
exhibits a small blueshift. Its line profile 
has been described by a raytracing model of a luminous disk in the Kerr metric.\cite{laor}  The Fe K$\alpha$ line provides evidence for
an accretion disk extending very close to the event horizon of the black hole
(however, alternative
models exist, i.e. see Titarchuk in these proceedings). 
Although the origin of the nonthermal radiation bathing the disk is not fully understood, it is likely produced by inverse Compton scattering
of $\sim 100$~keV electrons with thermal UV radiation produced 
by the disk. 

The relativistically-broadened 
emission lines of C~VI, N~VII, and O~VIII Ly$\alpha$ have been tentatively
identified in a handful of active galaxies.\cite{branduardi}
This identification has been controversial since the same X-ray spectra
can be interpreted with photoelectric absorption features
produced by plasma that is known to exist several parsecs 
away from the black hole.\cite{lee}
Our primary goal is to produce a theoretical prediction for
such a recombination spectrum, accounting for relativistic effects.
Since the raytrace calculation by Laor\cite{laor} omitted the 
physics of the line emission mechanism in the disk, it 
introduced a power-law emissivity $\epsilon \propto r^{-q}$ 
as a function of radius $r$, 
with $q$ as a free parameter.
Another purpose of our work is to test whether or not this line 
emissivity index $q$ is directly
related to the emissivity index of the total radiative energy
of the disk, $q_{T}$, as a function of $r$.  
The Shakura \& Sunyaev\cite{ss73} disk model implies that
$q_{T} =3$, but in situations where the disk is heated by a coupling with the
Kerr hole, $q_{T} = 3.5$.\cite{agolk}  However, fits to the lines observed
with the {\it XMM-Newton} X-ray observatory show emissivity indices for 
Fe K$\alpha$ of $q =4.3$--5.0\cite{wilms}, that are 
larger than the theoretical predictions of $q_T$.
Could $q \neq q_T$ be due to 
atmospheric effects in the accretion disk?
We test this hypothesis for the fluorescence and 
recombination X-ray lines of the most abundant elements. 

\section{Disk Atmosphere Structure}
We will describe how we calculate the disk atmosphere structure, with the
understanding that we are forced to introduce simplifying assumptions,
specifically due to our ignorance of the dynamical processes 
responsible for electron acceleration in the atmosphere and disk corona.
Underlying the atmosphere, we assume an accretion disk interior 
model,\cite{ss73} modified by the general relativistic corrections 
applicable in the Kerr metric.\cite{thorne}  We build upon this model by 
calculating the ionization structure of the
atmospheric interface between a fully ionized corona
and the disk interior, with the atmosphere 
energized by external nonthermal irradiation
from above and blackbody radiation from below.
We do not know the relationship between the energy 
(per unit disk area) dissipated 
in the disk $U_T=U_T(r)$, the energy emitted as blackbody radiation $U_{bb}$,
and the nonthermal radiation energy $U_{nt}$, so we choose
$U_T = U_{bb} + U_{nt}$, and fix $U_{bb} = U_{nt}$ 
for all $r$. We also assume that 
the spectral index of this nonthermal radiation (which is an observable)
is constant with radius, but that its low-energy cutoff is related to the 
local temperature of the disk interior,
consistent with a Comptonization origin.  

\begin{figure}[hp] 
\centerline{\epsfxsize=5.0in\epsfbox{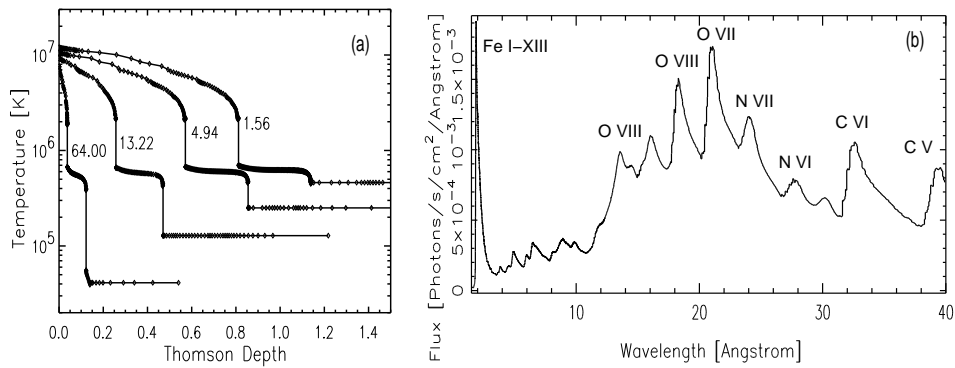}}
\caption{(a) Atmospheric temperature as a function of vertical Thomson depth. 
Selected radii in units of gravitational radius are shown. 
(b) X-ray line spectrum of the disk atmosphere
at an inclination of 30$^\circ$, without the
continuum emission, and including most relativistic effects. \label{fig:tau}}
\end{figure}

Our calculation relies on the dominance of photoionization
and recombination in determining the temperature and ionization state
of the plasma in the accretion disk atmosphere, 
such that collisional ionization is unimportant.
As the depth increases, the atmosphere 
becomes optically thick in X-rays, finally reaching a
disk interior dominated by collisional ionization.
Discrete atmospheric layers are produced by 
the thermally stable phases of the photoionized plasma, 
as seen in Fig. \ref{fig:tau}a. The intermediate layer 
at temperature $T \sim 6 \times 10^5$~K, as well as the deepest layer, 
emit copious recombination emission in the soft X-ray band,
while the Fe K$\alpha$ fluorescence emission is produced only at 
the deepest layer, at $T \lesssim 4 \times 10^5$~K.
The Thomson depth of the X-ray atmosphere decreases with 
radius.  The temperature of the deepest layer is set to
the temperature of the disk interior. 

We choose model parameters applicable to an active galaxy such as MCG~-6-30-15, which shows a prominent broad Fe K$\alpha$ line, and also the putative soft X-ray broad lines.  We set the accretion rate to $\dot{M} = 10^{24}$~g~s$^{-1}$, the power-law continuum photon index to $\Gamma = 2.1$, the high-energy exponential cutoff of the hard X-ray spectrum to $E_{\rm cut} = 150$~keV, the black hole mass to $M = 10^7$~\Msun, and the dimensionless black hole spin to $a = 0.998$. The inner disk torque is set to a nonzero value, increasing the accretion efficiency by 0.08.  We included the general relativistic correction for the tidal force in the vertical direction, but we did not include the temperature corrections (which are affected by the torque at the inner boundary in any case). We binned the accretion disk to follow the Laor\cite{laor} prescription of $r_n = 1600 n^{-2} r_g$ with $n=2,3,...,36$, in $r_g= GM/c^2$ gravitational radii. The maximum number of vertical bins is 500, but 
200--300 is typical.

\section{Disk Atmosphere Spectrum}
We calculated the atomic X-ray spectrum with semi-analytic and with 
Monte Carlo methods (to improve the accuracy of the transfer calculation
and Compton scattering).  The spectra include recombination emission from the hydrogen-like and helium-like ions of C, N, O, Ne, Mg, Si, S, Ar, Ca, and Fe, plus the L-shell ions of Fe, and H and He. We use a fluorescence yield of 0.34 for all Fe ions with an M-shell electron. Fe ions with just L-shell electrons are assumed to have zero fluorescence yield, since photon trapping will suppress fluorescence.\cite{ross} 
The semi-analytic X-ray spectrum shown in Fig. \ref{fig:tau}b exhibits a series of distinct sawtooth peaks, each due to a line broadened by Doppler shifts and gravitational redshift. The most prominent lines are O~VIII, O~VII, N~VII, C~VI, and C~V.  There is {\it not} a strong contribution to the emission
from Fe L-shell ions, as reported by other authors.\cite{balmodel}
The equivalent width relative to the
incident continuum of O~VIII Ly$\alpha$ is 23~eV, while that of the entire line complex, from 13 to 35~\AA, is 110~eV.  The latter is well within the detection limit of the {\it Chandra} and {\it XMM-Newton} X-ray observatories.  

One candidate for this kind of emission is NGC~4051, which exhibits a soft X-ray flux excess that can be fit by an O~VIII disk line complex.\cite{ogle}  
However, the sawtooth profiles of the emission lines
are not clear in the data.
Both MCG~-6-30-15 and Mrk~766 could exhibit such lines, but our Monte Carlo\cite{mauche} has not yet reproduced the bright N VII line and absent Lyman series. Also, while the radiative recombination continua are reduced in flux in the model, they are absent in the 
observed\cite{branduardi} spectral fits.
We will soon include line opacities to model the potential line splitting.

\section{The Line Emissivities}
The modeled emissivities as a function of radius of 
O~VII, O~VIII, N~VII, N~VI, C~VI, and C~V, 
all track the input radiative energy index $q_T$ to 
within $\Delta q = \pm 0.5$. The hydrogen-like ion lines have systematically larger values of $q$ than the helium-like ion lines.  For the Fe K$\alpha$ line at 6.4~keV, $q = 3.0$, unless the disk interior $T \gtrsim 4 \times 10^5$~K, in which
case fluorescence is suppressed.  
However, for both Fe~XXVI and Fe~XXV Ly$\alpha$ lines at 6.7~keV and 6.9~keV, $q = 4.0$, which is larger than the $q_T = 3.0$ input.  This steep emissivity $q$ is due to a thickening of the hot atmosphere and an increase in density at small radii.  Atmospheric effects increase the emissivity indices significantly only for Fe XXV and Fe XXVI lines. The models show that a power-law description of the line emissivities is, in general, only 
a rough approximation.

\end{document}